\documentclass[english]{article}
\usepackage{lmodern}

\usepackage[T1]{fontenc}
\usepackage[latin9]{inputenc}
\usepackage{booktabs}
\usepackage{mathrsfs}
\usepackage{amsmath}
\usepackage{graphicx}

\makeatletter
\newcommand{\lyxaddress}[1]{
\par {\raggedright #1
\vspace{1.4em}
\noindent\par}
}

\usepackage{mathrsfs}   
\usepackage{slashed}     
\usepackage{bbold}  
\usepackage{url}
\usepackage{graphicx}
\usepackage[colorlinks=true,linkcolor=redLinks,citecolor=greenLinks,urlcolor=redLinks, pdfborder={0 0 1}]{hyperref}
\usepackage{xcolor}
\usepackage{framed}
\usepackage[numbers,sort&compress]{natbib}
\usepackage{amsmath}

\allowdisplaybreaks

\colorlet{shadecolor}{gray!15}

\definecolor{greenLinks}{rgb}{0, 0.6, 0} 
\definecolor{blueLinks}{rgb}{0, 0, 0.6}
\definecolor{redLinks}{rgb}{0.6, 0, 0}
\definecolor{tempText}{rgb}{0.55, 0.10,0.67}
\definecolor{eprintLinks}{rgb}{0.4, 0.4, 0.4}
\definecolor{journalLinks}{rgb}{0.6, 0, 0}

\newcommand{\MYhref}[3][redLinks]{\href{#2}{\color{#1}{#3}}}%

\def\gsim{\raise0.3ex\hbox{$\;>$\kern-0.75em\raise-1.1ex\hbox{$\sim\;$}}}
\def\lsim{\raise0.3ex\hbox{$\;<$\kern-0.75em\raise-1.1ex\hbox{$\sim\;$}}}
\def\znbb{0\nu\beta\beta}
\def\meff{\langle m_{\nu} \rangle}
\def\pslash{\slashed{p}}

\usepackage{multirow}
\let\orig@Hy@EveryPageAnchor\Hy@EveryPageAnchor
\def\Hy@EveryPageAnchor{%
    \begingroup
    \hypersetup{pdfview=Fit}%
    \orig@Hy@EveryPageAnchor
    \endgroup
}

\let\oldFootnote\footnote
\newcommand\nextToken\relax

\renewcommand\footnote[1]{%
    \oldFootnote{#1}\futurelet\nextToken\isFootnote}

\newcommand\isFootnote{%
    \ifx\footnote\nextToken\textsuperscript{,}\fi}

\makeatother

\usepackage{babel}
\begin{document}

\title{{\Large{}\vspace{-1.0cm}} \hfill {\normalsize{}IFIC/16-50} 
\\*[10mm] Gauge vectors and double beta decay{\Large{}\vspace{0.5cm}}}

\author{{\Large{}Renato M. Fonseca}\thanks{E-mail: renato.fonseca@ific.uv.es} 
\ and {\Large{}Martin Hirsch}\thanks{E-mail: mahirsch@ific.uv.es} 
\date{}}

\maketitle

\lyxaddress{\begin{center}
{\Large{}\vspace{-0.5cm}}\href{http://www.astroparticles.es/}{AHEP Group},
\href{http://webific.ific.uv.es/web/en}{Instituto de Física Corpuscular},
\href{http://www.csic.es}{C.S.I.C.}/\href{http://www.uv.es/}{Universitat de València}\\
Parc Científic de Paterna.  Calle Catedrático José Beltrán, 2 E-46980
Paterna (Valencia) -- Spain
\par\end{center}}

\begin{center}
\today
\par\end{center}
\begin{abstract}
We discuss contributions to neutrinoless double beta
($0\nu\beta\beta$) decay involving vector bosons. The starting point
is a list of all possible vector representations that may contribute
to $0\nu\beta\beta$ decay via $d=9$ or $d=11$ operators at tree level. We then
identify gauge groups which contain these vectors in the adjoint
representation.  Even though the complete list of vector fields that
can contribute to $0\nu\beta\beta$ up to $d=11$ is large (a total of
46 vectors), only a few of them can be gauge bosons of
phenomenologically realistic groups.  These latter cases are discussed
in some more detail, and lower (upper) limits on gauge boson masses
(mixing angles) are derived from the absence of $0\nu\beta\beta$ decay.

\end{abstract}
\pagebreak{}

\section{\label{sec:1}Introduction}

Neutrinoless double beta decay ($\znbb$ decay) is the most sensitive
experimental probe of lepton number violating (LNV) extensions of the
Standard Model (SM) --- for a review see for instance
\cite{Deppisch:2012nb}.  The latest experimental half-life bounds
\cite{KamLAND-Zen:2016pfg,Agostini:2013mzu,Agostini:2016zzz}
correspond to upper limits on the effective Majorana mass of the
neutrino of the order of $\meff \lsim (0.1-0.2)$ eV, depending on
nuclear matrix elements \cite{Muto:1989cd,Faessler:2012ku}.  However,
from the theoretical point of view the mass mechanism represents only
one out of many possible contributions to the $\znbb$ decay
amplitude. In fact, quite a number of papers on $\znbb$ decay
involving exotic (\textit{i.e.} beyond SM) scalars can be found in the
literature; see
\cite{Mohapatra:1986su,Hirsch:1995ek,Hirsch:1995cg,Hirsch:1996ye,%
	Pas:1998nn,Choubey:2012ux,Gu:2011ak,Kohda:2012sr,delAguila:2012nu} for some examples.

On the other hand, there are very few publications discussing the
contribution of exotic vector fields to this process.  The best-known
example of $\znbb$ decay induced by an exotic vector is perhaps the
charged boson $W_R$ exchange in the context of left-right
symmetric models. To the best of our knowledge, $\znbb$ decay in these models
was discussed for the first time in \cite{Mohapatra:1980yp}. This
subject has been studied in detail in many papers since
then. Contributions to $\znbb$ decay from vector lepto-quarks
were studied in \cite{Hirsch:1996ye}. A list of potential exotic
vector contributions to the long-range part of the $\znbb$ decay
amplitude \cite{Pas:1999fc} has been given in \cite{Helo:2016vsi}.
And, finally, there are a few papers on $\znbb$ decay in models based
on the group $SU(3)_c\times SU(3)_L\times U(1)_X$ (331, for short)
\cite{Pleitez:1993gc,Montero:2000ar}.

A general decomposition of $d=9$ contributions to $\znbb$ decay has
been derived in \cite{Bonnet:2012kh}. However, that paper focuses on
scalars and does not discuss exotic vectors in detail.\footnote{A
	table with the quantum number of vectors under $SU(3)_C\times
	U(1)_{\rm E.M.}$ is shown in \cite{Bonnet:2012kh}.}  To fill this
gap, in this paper we study systematically all possible exotic vectors
that give tree-level contributions to $\znbb$ decay (associated to 
$d=9$ and $d=11$ operators, as discussed latter).

Let us start by recalling some basic (but important) aspects of
$\znbb$ decay.  At tree-level a non-zero $\znbb$ decay amplitude can
be generated at $d=9$ by only two topologies (I and II)
\cite{Bonnet:2012kh} --- see figure \ref{fig:topos}. The internal
bosons can either be scalars or vectors; for example, in topology I one might
have in the internal lines the combination vector--fermion--vector
($V\psi V$), vector--fermion--scalar ($V\psi S$) or
scalar--fermion--scalar ($S\psi S$).  Note, however, that for topology II not
all combinations are equally important. This is because diagrams with
$VSS$ and $VVV$ contain a derivative and thus, effectively these
contributions are proportional to $p_F/\Lambda_{LNV}^6$, compared to
$1/\Lambda_{LNV}^5$ for the other combinations. Here $p_F$ is the
Fermi momentum of the nucleons, of the order of $(100-200)$ MeV, and
$\Lambda_{LNV}$ is the scale of lepton number violation, which is also
associated to the mass of the exotic particles.

All topology II diagrams contribute only to the so-called short-range
amplitude of $\znbb$ decay \cite{Pas:2000vn}. This part of the
amplitude involves diagrams in which all virtual particles have masses
larger than $p_F$. In topology I, however, it is possible that the internal
fermion $\psi$ is a light neutrino. In this case, one talks about a
long-range contribution \cite{Pas:1999fc}, since $p_F$ corresponds to
larger, inter-nucleon distances (typically of the order of a
femtometer). This distinction between short- and long-range amplitudes is
important, since the nucleon hard-core strongly suppresses the matrix
elements of the short-range part of the amplitude.

\begin{center}
	\begin{figure}[tbph]
		\begin{centering}
			\includegraphics[scale=0.8]{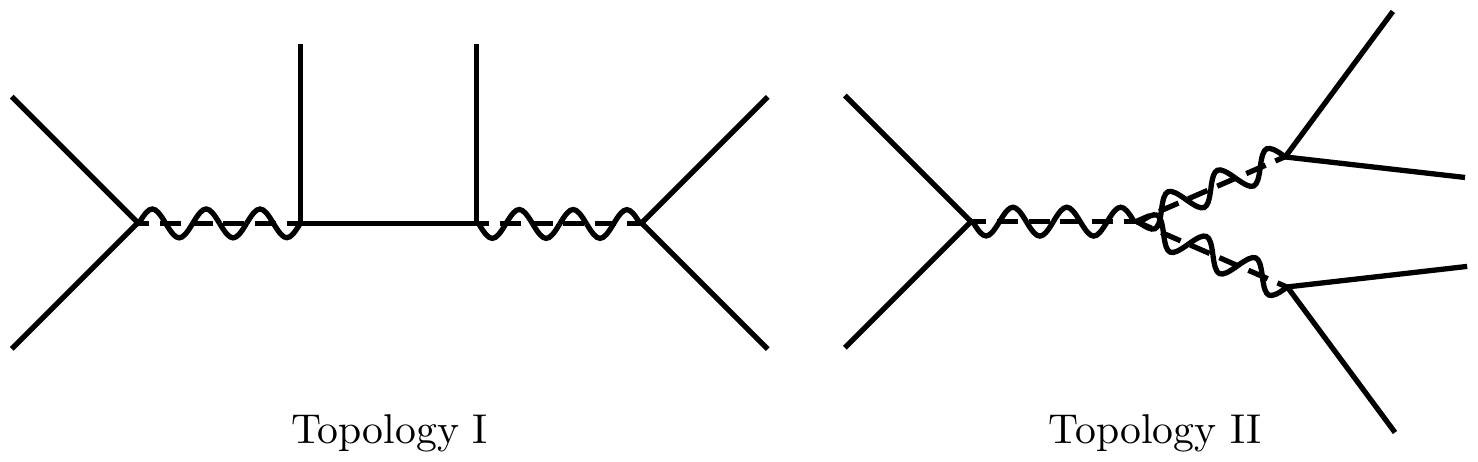}
			\par
		\end{centering}
		\protect\caption{\label{fig:topos}
			Tree-level topologies for the $\znbb$ decay operator at $d=9$. 
			Internal bosons can be scalars or vectors.}
	\end{figure}
	\par
\end{center}

Within the Standard Model there is only one non-renormalizable
dimension 5 operator: the famous Weinberg operator $O_W =
LL HH$ \cite{Weinberg:1979sa}.  However, at higher
order, many more $\Delta L=2$ operators can be written down ---
indeed, the complete list of $\Delta L=2$ operators with no
derivatives/gauge bosons was given in \cite{Babu:2001ex}. Three $d=7$,
nine $d=9$ and thirteen $d=11$ operators contribute to $\znbb$ decay
at tree-level.\footnote{ The $d=7$ and three of $d=9$ $\Delta L=2$
	operators must be complemented by a SM charged-current interaction,
	in order to generate a $\znbb$ decay diagram.}  From this effective
operator point of view, for example the mass mechanism is simply
described by ${\cal
	O}_{47}=\overline{L}\overline{L}\overline{Q}\overline{Q}QQH^{*}H^{*}
$. (The number ``47'' corresponds to the notation of \cite{Babu:2001ex}.)
If we contract this operator as $({\overline{Q}}Q)(\overline{L}) (H^*
H^*)(\overline{L})({\overline{Q}}Q)$, the internal particles of the
diagram are fixed to be $V=\left(\mathbf{1},\mathbf{3},0\right)
\equiv W$ and
$\psi=\left(\mathbf{1},\mathbf{2},-1/2\right)
\equiv L$. For the mass mechanism diagram, one
just needs to select the isospin component of the external fermions to be
$({\overline u}d) (e)(e)({\overline u}d)$, and connect the internal
neutrinos via the Weinberg operator as shown in figure
\ref{fig:massMech}.

\begin{center}
	\begin{figure}[tbph]
		\begin{centering}
			\includegraphics[scale=0.8]{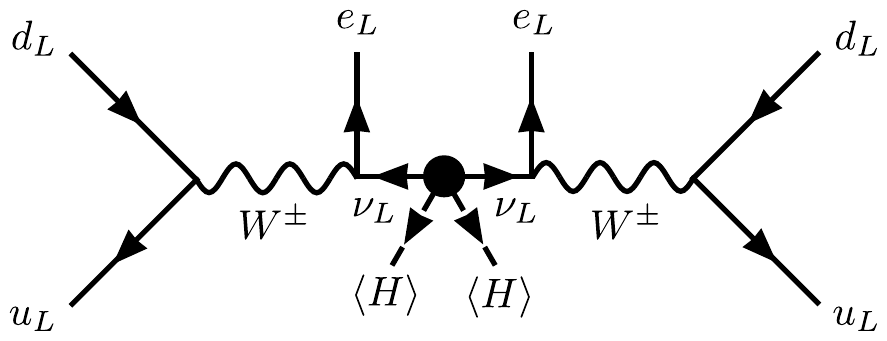} 
			\par
		\end{centering}
		\protect\caption{\label{fig:massMech} Mass mechanism of neutrinoless double beta
			decay. In SM invariant language, this corresponds to the operator
			${\cal O}_{47}$ \cite{Babu:2001ex} contracted as $({\overline{Q}}Q)(\overline{L})
			(H^* H^*)(\overline{L})({\overline{Q}}Q)$.}
	\end{figure}
	\par\end{center}

It is possible to find then all vectors contributing potentially to
$\znbb$ decay by decomposing all the relevant $d=7,9,11$ operators in
a similar way. This is done in section \ref{sec:2}. However, all
experimentally observed fundamental vectors are gauge vectors so, in a
second step, we search for gauge groups which contain one of our
exotic vectors in the adjoint representation. Apart from the mass
mechanism (\textit{i.e.} the SM vectors), we recover the vectors of
the left-right symmetric group and the 331 group mentioned above. In
addition to these possibilities, we find more exotic ones, such as the
vectors of the Pati-Salam group, $SU(5)$ or more peculiar groups, such
as for example $Sp(8)\times U(1)^n$ and $F_4\times U(1)^n$.  However,
most of these groups are not phenomenologically interesting for
$\znbb$ decay, either because (i) their vectors lead to proton decay
or (ii) it is not possible to construct viable models with the SM
fermion content. This is discussed in section \ref{sec:2} as well. In
section \ref{sec:Pheno} we then analyze the constraints on the (few) viable
exotic vectors potentially contributing to $\znbb$ decay. 
We then finish with a short summary.

\section{\label{sec:2}Vector contributions to $\znbb$ decay}

\subsection{$\Delta L=2$ operators and vectors}

Both in topology I and in topology II internal vector fields couple to
a pair of external fermion lines (\textit{i.e.}, leptons and/or quarks),
hence their gauge quantum numbers are not arbitrary. In order to
explore this fact, take as a starting point the following list of 
dimension nine ($d=9$) $\znbb$ operators:
\begin{eqnarray}\label{eq:d9} 
\mathcal{O}^{\textrm{d=9}}=
\overline{Q}\overline{Q}\overline{L}\overline{L}\overline{d^{c}}\overline{d^{c}},
QQu^{c}u^{c}\overline{L}\overline{L}, 
\overline{Q}Qu^{c}\overline{d^{c}}\overline{L}\overline{L},
\overline{Q}\overline{d^{c}}\overline{d^{c}}u^{c}\overline{L}e^{c},
Qu^{c}u^{c}\overline{d^{c}}\overline{L}e^{c}, 
u^{c}u^{c}\overline{d^{c}}\overline{d^{c}}e^{c}e^{c}.
\end{eqnarray}
A vector field $V_{\mu}$ must couple to a left and a right-handed
fermion or, equivalently, to a combination $X\overline{Y}$ where both
$X$ and $Y$ are left-handed fields. Given the above list of $d=9$
operators it is straightforward to find all possible fermion bilinears
coupling to vectors and identify the potentially interesting vector
representations:
\begin{align}
V_{\mu}^{\textrm{d=9}}= & \left(\mathbf{8},\mathbf{3},0\right),\left(\mathbf{8},\mathbf{1},0\right),\left(\mathbf{1},\mathbf{3},0\right),\left(\mathbf{1},\mathbf{1},0\right),\left(\mathbf{6},\mathbf{2},\frac{1}{6}\right),\left(\mathbf{3},\mathbf{2},\frac{1}{6}\right),\left(\mathbf{3},\mathbf{3},\frac{2}{3}\right),\nonumber \\
& \left(\mathbf{3},\mathbf{1},\frac{2}{3}\right),\left(\overline{\mathbf{6}},\mathbf{2},\frac{5}{6}\right),\left(\mathbf{3},\mathbf{2},-\frac{5}{6}\right),\left(\mathbf{8},\mathbf{1},1\right),\left(\mathbf{1},\mathbf{1},1\right),\left(\mathbf{1},\mathbf{2},-\frac{3}{2}\right)\,.\label{eq:dim9reps}
\end{align}
However, before exploring this list, it is worth reminding that the neutrino mass contribution to $\znbb$ decay actually comes from a $d=11$ effective operator with two Higgs
fields. Hence, in the search for new contributions to $\znbb$ one 
should not stop at the level of $d=9$ operators (see section \ref{sec:1}). At $d=11$ one 
finds thirteen operators that contribute to $\znbb$: 
\begin{align}
\mathcal{O}^{\textrm{d=11}}=\; & u^{c}u^{c}\overline{d^{c}}\overline{d^{c}}e^{c}e^{c}HH^{*},\,u^{c}u^{c}\overline{d^{c}}\overline{d^{c}}\overline{L}\overline{L}HH,\,u^{c}u^{c}\overline{d^{c}}Qe^{c}\overline{L}HH^{*},\,u^{c}u^{c}QQe^{c}e^{c}H^{*}H^{*},\,u^{c}u^{c}QQ\overline{L}\overline{L}HH^{*},\nonumber \\
& u^{c}\overline{d^{c}}\overline{d^{c}}\overline{Q}e^{c}\overline{L}HH^{*},\,u^{c}\overline{d^{c}}\overline{Q}Qe^{c}e^{c}H^{*}H^{*},\,u^{c}\overline{d^{c}}\overline{Q}Q\overline{L}\overline{L}HH^{*},\,u^{c}\overline{Q}QQe^{c}\overline{L}H^{*}H^{*},\,\overline{d^{c}}\overline{d^{c}}\overline{Q}\overline{Q}e^{c}e^{c}H^{*}H^{*},\nonumber \\
& \overline{d^{c}}\overline{d^{c}}\overline{Q}\overline{Q}\overline{L}\overline{L}HH^{*},\,\overline{d^{c}}\overline{Q}\overline{Q}Qe^{c}\overline{L}H^{*}H^{*},\,\overline{Q}\overline{Q}QQ\overline{L}\overline{L}H^{*}H^{*}\,.
\end{align}

\begin{figure}[tbph]
	\begin{centering}
		\includegraphics[scale=0.8]{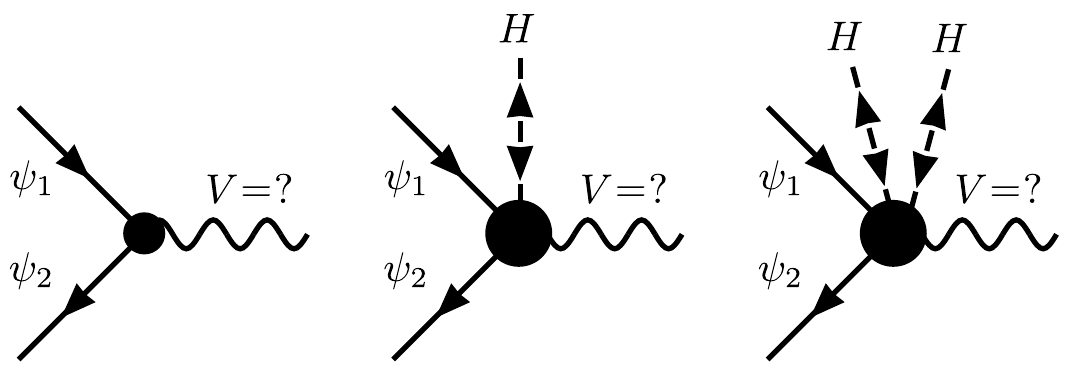}
		\par\end{centering} \protect\caption{\label{fig:VPsiPsi_vertex} Vector
		fields $V$ couple directly to a fermion bilinear in the $d=9$
		operators (left diagram). For the $d=11$ operators one must also
		take into account up to two additional Higgs field insertions while
		searching for possible vectors (center and right diagrams).}
\end{figure}

\noindent
Allowing for up to two external Higgs fields, the vectors can not only
couple to a $X\overline{Y}$ fermion combination, but also effectively
have a coupling to $X\overline{Y}H^{(*)}$ or
$X\overline{Y}H^{(*)}H^{(*)}$, where $X$ and $Y$ are still left-handed
fields and $H^{(*)}$ represents the Higgs field (possibly conjugated);
see figure \ref{fig:VPsiPsi_vertex}.

Taking into account these additional Higgs insertions, one obtains the
following larger list of potentially interesting vector
representations:
\begin{align}
V_{\mu}^{\textrm{dim-11}} & =\left(\left[\mathbf{1},\mathbf{8}\right],\left[\mathbf{1},\mathbf{3}\right],\left[0,1,2\right]\right),\left(\mathbf{1},\left[\mathbf{2},\mathbf{4}\right],\left[-\frac{1}{2},\frac{3}{2}\right]\right),\left(\left[\mathbf{1},\mathbf{8}\right],\mathbf{5},\left[0,1\right]\right),\left(\mathbf{8},\mathbf{2},\left[\frac{1}{2},\frac{3}{2}\right]\right),\left(\mathbf{8},\mathbf{4},-\frac{1}{2}\right)\,,\nonumber \\
& \left(\left[\mathbf{3},\overline{\mathbf{6}}\right],\left[\mathbf{1},\mathbf{3}\right],\left[-\frac{4}{3},-\frac{1}{3},\frac{2}{3}\right]\right),\left(\left[\mathbf{3},\overline{\mathbf{6}}\right],\left[\mathbf{2},\mathbf{4}\right],\left[-\frac{11}{6},-\frac{5}{6},\frac{1}{6},\frac{7}{6}\right]\right),\left(\mathbf{3},\mathbf{5},\left[-\frac{1}{3},\frac{2}{3}\right]\right)\,.\label{eq:dim11reps}
\end{align}
Square brackets in this list represent independent alternatives for
each quantum number --- hence
($[\mathbf{1},\mathbf{8}]$, $[\mathbf{1},\mathbf{3}]$, $[0,1,2]$) stands
for 12 different representations, for example.  In total, there are 53
representations listed in equation (\ref{eq:dim11reps}).

\begin{figure}[tbph]
	\begin{centering}
		\includegraphics[scale=0.8]{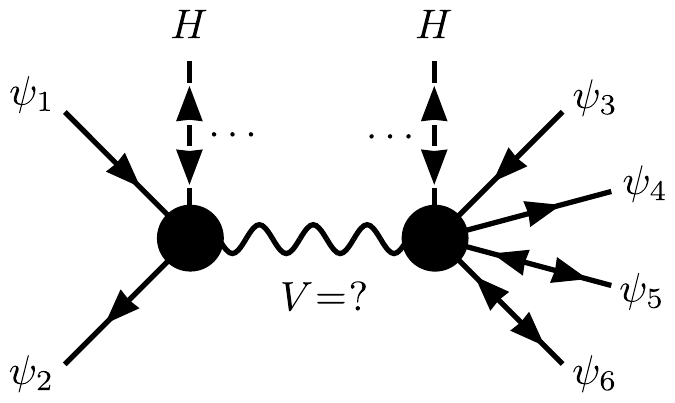}
		\par\end{centering}
	
	\protect\caption{\label{fig:VPsiPsi_vertex_plus_rest}There are several
		dimension nine and eleven operators of the form
		$X_{1}\overline{X_{2}}X_{3}\overline{X_{4}}X_{5}X_{6}$ or
		$X_{1}\overline{X_{2}}X_{3}\overline{X_{4}}\overline{X_{5}}\overline{X_{6}}$
		plus Higgses. One must then ensure that each vector field $V$
		has the adequate quantum numbers to couple both to a combination
		$X_{1}\overline{X_{2}}$ (plus, possibly, some Higgses), and at the
		same time to a combination $X_{3}\overline{X_{4}}X_{5}X_{6}$ or
		$X_{3}\overline{X_{4}}\overline{X_{5}}\overline{X_{6}}$ (again, plus
		Higgses). }
\end{figure}

However, not all 53 vectors constructed in this way do actually
contribute to $\znbb$ decay. First of all, neither
$(\mathbf{1},\mathbf{1},0)$ nor $(\mathbf{8},\mathbf{1},0)$ carry
electric charge and thus they can be trivially excluded. Furthermore, one
actually has to ensure that the diagram is completable on the
other side of the vector boson line (see figure
\ref{fig:VPsiPsi_vertex_plus_rest}). This requirement, in turn,
excludes the representations $(\mathbf{1},\mathbf{4},\frac{3}{2})$,
$(\mathbf{1},\mathbf{5},0)$, $(\mathbf{3},\mathbf{4},-\frac{5}{6})$,
$(\mathbf{3},\mathbf{5},\frac{2}{3})$ and $(\mathbf{8},\mathbf{5},0)$.
We are then left with 46 potentially interesting vector
representations. All of them can play a role in either topology I or II,
provided that one completes the diagrams appropriately (this might require
adding two Higgs insertions). We do not provide here a list of all such valid
configurations, since these can be easily constructed.

\subsection{Gauge groups}

Given the list of vectors discussed above, we now turn to the search
for possible gauge groups (see also \cite{Biggio:2016wyy}). These groups must contain the SM group and
for any vector to be a gauge boson of such a group, it must form a
sub-representation of the adjoint representation of the larger group.
There are several subtleties involved in this search, hence we shall
start by making a few remarks.  Firstly, a given group may break
into a subgroup in more than one way. For example, $SU(3)$ contains
$SU(2)$ but we may either have the branching rule
$\mathbf{3}\rightarrow\mathbf{2}+\mathbf{1}$ or
$\mathbf{3}\rightarrow\mathbf{3}$. Mathematically both represent valid
embeddings of $SU(2)$ in $SU(3)$, yet this does not mean that they are
equally useful for model building. In particular, if we consider the
Standard Model $SU(2)_{L}$ group, the embedding corresponding to the
branching rule $\mathbf{3}\rightarrow\mathbf{3}$ can be ruled out
since no matter which $SU(3)$ representation is chosen, it is
impossible to obtain $SU(2)_{L}$ doublets.\footnote{The reason is the
	following. Any $SU(3)$ representation will be picked up in the
	product of $n$ fundamental representations of $SU(3)$, for a
	sufficiently large $n$. As such, if the fundamental of $SU(3)$
	branches into an $SU(2)_{L}$ triplet, an $SU(3)$ representation can
	only branch into $SU(2)_{L}$ representations contained in the
	product of $SU(2)_{L}$ triplets (\textit{i.e.},
	$\mathbf{1},\mathbf{3},\mathbf{5},\cdots$ representations, but never 
	$\mathbf{2},\mathbf{4},\mathbf{6},\cdots$).}

Another important issue concerns the normalization of $U(1)$ charges.
Consider a group $G$ with an $SU(3)_{C}\times SU(2)_{L}\times
U(1)^{n}$ subgroup, $n>0$. Without writing down the complete model
based on $G$, it is not possible to determine how the $n$ $U(1)$'s
combine to form the $U(1)_{Y}$ hypercharge group. Take, for example,
the $SU(3)_{C}\times SU(2)_{L}\times SU(2)$ group: the adjoint
representation contains the gluons
$\left(\mathbf{8},\mathbf{1},\mathbf{0}\right)$, the $W$'s
$\left(\mathbf{1},\mathbf{3},\mathbf{0}\right)$ and
$\left(\mathbf{1},\mathbf{1},\mathbf{3}\right)$.  If we break the
$SU(2)$ group to $U(1)$, this last representation branches into
$\left(\mathbf{1},\mathbf{1},0\right)+\left(\mathbf{1},\mathbf{1},\pm1\right)$,
yet one is \textit{a priori} free to consider that the hypercharge is
a multiple of the last quantum number. Hence, without first trying to
build a complete model --- including all relevant fermions and scalars ---
one has to consider that any SM representation of the form
$\left(\mathbf{1},\mathbf{1},y\right)$ is potentially contained in the
adjoint of $SU(3)_{C}\times SU(2)_{L}\times SU(2)$.  Furthermore, if
necessary, one can add extra $U(1)$'s to the original group $G$ in
order to build the SM $U(1)_{Y}$ hypercharge group (\textit{i.e.}, instead of
$G$ one might consider $G\times U(1)^{m}$). This is indeed what
happens in the well-known models based on the $SU(3)_{C}\times
SU(3)\times U(1)$ and $SU(3)_{C}\times SU(2)_{L}\times SU(2)\times
U(1)$ groups.

For these reasons, we will temporarily (a) ignore whether or not a
full viable model can be made out of particular groups/embeddings, and
(b) ignore $U(1)$'s altogether (since all that matters, according to the
previous discussion, is whether or not the hypercharge of a vector
field is non-zero). We can then try to embed the SM group in a
simple group $G$, $SU(3)_{C}\times G$, or maybe $SU(2)_{L}\times
G$.\footnote{We stress again that eventual extra $U(1)$'s, needed to
	correctly form the hypercharge group, are being ignored at this
	stage.} A search was performed over all such groups where $G$ is
both simple and has at most 9 diagonal generators. This includes all
exceptional groups, all $SU(N<11)$, all $SO(N<20)$, and the $Sp(2N)$
groups up to $N=9$. Table \ref{tab:Groups_for_each_rep} contains the
results of this search. 

\begin{center}
	\begin{table}[tbh]
		\begin{centering}
			\begin{tabular}{cc}
				\toprule 
				Vector representation(s) & Minimal group(s) (without $U(1)$'s)\tabularnewline
				\midrule
				$(\mathbf{1},\mathbf{1},$\framebox[1.1\width]{$y=1$}$,2)$ & \framebox[1.1\width]{$SU(3)_{C}\times SU(2)_{L}\times SU(2)$}\tabularnewline
				\framebox[1.1\width]{$\left(\mathbf{1},\mathbf{2},y=\frac{1}{2},\frac{3}{2}\right)$} & \framebox[1.1\width]{$SU(3)_{C}\times SU(3)$}, $SU(3)_{C}\times Sp(4)$\tabularnewline
				\framebox[1.1\width]{$\left(\mathbf{1},\mathbf{3},0\right)$} & \framebox[1.1\width]{$SU(3)_{C}\times SU(2)_{L}$}\tabularnewline
				$\left(\mathbf{1},\mathbf{3},y=1,2\right)$ & $SU(3)_{C}\times Sp(4)$\tabularnewline
				$\left(\mathbf{1},\mathbf{4},y=\frac{1}{2}\right)$ & $SU(3)_{C}\times SU(5)$, $SU(3)_{C}\times Sp(6)$, $SU(3)_{C}\times G_{2}$\tabularnewline
				$\left(\mathbf{1},\mathbf{5},1\right)$ & $SU(3)_{C}\times SO(7)$, $SU(3)_{C}\times Sp(6)$\tabularnewline
				$(\mathbf{3},\mathbf{1},y=-\frac{4}{3},-\frac{1}{3},$\framebox[1.1\width]{$\frac{2}{3}$}$)$ & \framebox[1.1\width]{$SU(4)\times SU(2)_{L}$}\tabularnewline
				$\left(\mathbf{3},\mathbf{2},y=-\frac{11}{6},-\frac{5}{6},\frac{1}{6},\frac{7}{6}\right)$ & $SU(5)$, $Sp(8)$, $F_{4}$\tabularnewline
				$\left(\mathbf{3},\mathbf{3},y=-\frac{4}{3},-\frac{1}{3},\frac{2}{3}\right)$ & $SU(6)$, $SO(9)$\tabularnewline
				$\left(\mathbf{3},\mathbf{4},y=-\frac{11}{6},\frac{1}{6},\frac{7}{6}\right)$ & $SU(7)$, $Sp(10)$, $E_{6}$\tabularnewline
				$\left(\mathbf{3},\mathbf{5},y=-\frac{1}{3}\right)$ & $SU(8)$, $SO(11)$\tabularnewline
				$\left(\mathbf{6},\mathbf{1},y=-\frac{2}{3},\frac{1}{3},\frac{4}{3}\right)$ & $Sp(6)\times SU(2)_{L}$\tabularnewline
				$\left(\mathbf{6},\mathbf{2},y=-\frac{7}{6},-\frac{1}{6},\frac{5}{6},\frac{11}{6}\right)$ & $SU(8)$, $Sp(14)$, $F_{4}$\tabularnewline
				$\left(\mathbf{6},\mathbf{3},y=-\frac{2}{3},\frac{1}{3},\frac{4}{3}\right)$ & $SU(9)$, $SO(15)$, $Sp(12)$\tabularnewline
				$\left(\mathbf{6},\mathbf{4},y=-\frac{7}{6},-\frac{1}{6},\frac{5}{6},\frac{11}{6}\right)$ & $Sp(16)$ {[}{*}{]}\tabularnewline
				$\left(\mathbf{8},\mathbf{1},y=1,2\right)$ & $SU(6)\times SU(2)_{L}$, $SO(10)\times SU(2)_{L}$\tabularnewline
				$\left(\mathbf{8},\mathbf{2},y=\frac{1}{2},\frac{3}{2}\right)$ & $SU(9)$, $SO(12)$, $E_{6}$\tabularnewline
				$\left(\mathbf{8},\mathbf{3},0\right)$ & $SU(6)$, $SO(11)$\tabularnewline
				$\left(\mathbf{8},\mathbf{3},y=1,2\right)$ & $SO(14)$ {[}{*}{]}\tabularnewline
				$\left(\mathbf{8},\mathbf{4},\frac{1}{2}\right)$ & $SO(16)$ {[}{*}{]}\tabularnewline
				$\left(\mathbf{8},\mathbf{5},1\right)$ & $SO(18)$ {[}{*}{]}\tabularnewline
				\bottomrule
			\end{tabular}
			\par\end{centering}
		
		\protect\caption{\label{tab:Groups_for_each_rep}Minimal groups (second
			column) containing in their adjoint representation the indicated
			Standard Model vector sub-representations (first column). The exact
			value of the hypercharge of each representation is irrelevant: it
			only matters whether or not it is zero (see text). The list of
			groups shown here is of the form $G$, $SU(3)_{C}\times G$, or
			$SU(2)_{L}\times G$ where $G$ is a simple group with at most 9
			diagonal generators. Note, however, that it is straightforward to
			show that, except for those cases marked with an asterisk, the
			results are unchanged if this constraint is removed. Only minimal
			groups are displayed, hence any other group containing those listed above
			will obviously also work. Just the cases marked by a box lead to
			phenomenologically viable models, for the reasons explained in the text.}
	\end{table}
	
	\par\end{center}

However, table \ref{tab:Groups_for_each_rep} contains many vector
representations and groups which, in practice, are either
phenomenologically uninteresting or should be discarded altogether,
since they can not lead to models which contain the SM particle
content. As an example, consider the vectors
$\left(\mathbf{3},\mathbf{2},-\frac{5}{6}\right)$ and
$\left(\mathbf{3},\mathbf{2},\frac{1}{6}\right)$. Both can couple to a quark-lepton
pair and also to a pair of
quarks. As such, the simultaneous presence of these interactions leads to proton
decay, which means that such fields have to be very heavy, with masses
$\sim 10^{15}$ GeV or larger.
Consequently, such fields have a negligible impact on $ \znbb$ decay
phenomenology.

An even more important consideration is that for
many of the vector representations indicated in table
\ref{tab:Groups_for_each_rep}, the group listed must contain the SM gauge group
in a non-conventional way. This makes it very challenging to
accommodate the known fermion fields. For example, $F_{4}\times
U(1)^{n}$ is an unsuitable group: $F_4$ is an exceptional Lie group
which only has real representations, hence it cannot account for the
SM chirality (even if one considers extra $U(1)$'s).

As a further example, consider the
$\left(\mathbf{1},\mathbf{3},y=1,2\right)$ representations which can
be obtained, in theory, from a model with a gauge group as small as
$SU(3)_{C}\times Sp(4)\times U(1)^{n}$ for some $n$. We shall now show
that the Standard Model fermions cannot be embedded in representations
of such a group. Under the relevant branching rules, we have the
following breaking of $Sp(4)$ representations into sub-representations
of $SU(2)_{L}\times U(1)_{X}$:
\begin{align}
\mathbf{4} & \rightarrow\left(\mathbf{2},\pm1\right)\,,\label{eq:sp4_eq1}\\
\mathbf{5} & \rightarrow\left(\mathbf{1},\pm2\right)+\left(\mathbf{3},0\right)\,,\\
\mathbf{10} & \rightarrow\left(\mathbf{1},0\right)+\left(\mathbf{3},0\right)+\left(\mathbf{3},\pm2\right)\,,\\
\mathbf{14} & \rightarrow\left(\mathbf{1},0\right)+\left(\mathbf{1},\pm4\right)+\left(\mathbf{3},\pm2\right)+\left(\mathbf{5},0\right)\,,\\
\mathbf{16} & \rightarrow\left(\mathbf{2},\pm1\right)+\left(\mathbf{2},\pm3\right)+\left(\mathbf{4},\pm1\right)\,,\label{eq:sp4_eq2}\\
& \cdots\nonumber 
\end{align}
Note that \textit{a priori} the hypercharge group $U(1)_{Y}$ can be a
linear combination of $U(1)_{X}$ and the other $n$ $U(1)$ factor
groups. Even so, one can rule out the possibility of embedding the SM
in a $SU(3)_{C}\times Sp(4)\times U(1)^{n}$ model with the above
branching rules by, for example, counting $SU(2)_{L}$ doublets.  The
argument is the following. Since all $Sp(4)$ representations are real,
they must break into real and/or pairs of complex conjugated
representations of $SU(2)_{L}\times U(1)$. In turn, this means that
one will always have an even number of $SU(2)_{L}$ doublets, because
it can be shown that there are no doublets with $X=0$ regardless of which
$Sp(4)$ representations are chosen,\footnote{This is true for
	the representations in equations \eqref{eq:sp4_eq1}--\eqref{eq:sp4_eq2},
	but also for larger ones.} so any model based on this particular
embedding of the SM group in $SU(3)_{C}\times Sp(4)\times U(1)^{n}$
will necessarily have an even number of $SU(2)_{L}$ doublets. In fact,
one can be more specific: the number of doublets with a given color
quantum number (singlet, triplet, anti-triplet, ...) will be
even. This, however, does not lead to a satisfactory fermion sector,
as we must have three quark doublets plus (eventually) pairs of vector
quarks, hence the required number of fermion representations which are
color triplet/anti-triplets and $SU(2)_{L}$ doublets must be odd.

~

We have argued that there are no realistic models based on the gauge group $SU(3)_{C}\times
Sp(4)\times U(1)^{n}$ which contain the vector
representations $\left(\mathbf{1},\mathbf{3},y=1,2\right)$. Furthermore, a model based on
$F_{4}\times U(1)^{n}$ will not work either. Yet these are just
illustrative examples of the challenge of building realistic models
where the vector representations in table
\ref{tab:Groups_for_each_rep} show up. In fact, the only
groups and vectors for which we were able to find viable models (and
which give potentially interesting contributions to $\znbb$) besides
the SM $W=\left(\mathbf{1},\mathbf{3},0\right)$ are:\footnote{
	There are $n$ vector representations with the quantum numbers $\left(\mathbf{1},\mathbf{3},0\right)$
	in models based on the gauge group $SU(3)_{C}\times SU(2)^{n}\times U(1)_{Y}$
	provided that the Standard Model $SU(2)_{L}$ corresponds to the subgroup
	$\left[SU(2)^{n}\right]_{\textrm{diag}}$ \cite{Li:1981nk}. These models have a viable
	fermion spectrum and indeed they have recently been explored as a
	possible explanation for meson anomalies (see \cite{Boucenna:2016wpr} and references
contained therein). However, apart from non-universal flavor effects,
the interactions of the heavier $W'^{\pm}$ in these models are similar
to those of the SM $W^{\pm}$, hence their contribution to neutrinoless
double beta decay will be subdominant.
}
\begin{itemize}
	\item $\left(\mathbf{1},\mathbf{1},1\right)$ associated to
	$SU(3)_{C}\times SU(2)_{L}\times SU(2)\times U(1)$,
	\item $\left(\mathbf{1},\mathbf{2},y=\frac{1}{2},\frac{3}{2}\right)$
	associated to $SU(3)_{C}\times SU(3)\times U(1)$,
	\item $\left(\mathbf{3},\mathbf{1},\frac{2}{3}\right)$ associated to
	$SU(4)\times SU(2)_{L}\times U(1)$,
\end{itemize}
or bigger groups containing these ones. In the next section we will look carefully into these cases.

\section{\label{sec:3}Phenomenologically viable cases}
\label{sec:Pheno}

\subsection{Brief comments on non-gauge vectors}
\label{subesc:NGV}

Before we turn to a discussion of the phenomenologically viable
groups, we briefly comment on  non-gauge vectors. One can, of
course, add exotic vectors to the SM particle content without
enlarging the gauge group. A classical example is electro-weak scale
lepto-quarks \cite{Buchmuller:1986zs, Davidson:1993qk}.  Such
relatively light vectors can have interactions with the SM Higgs boson
\cite{Hirsch:1996qy}, which leads to a $\Delta L=2$ mixing between
different lepto-quarks states after electro-weak symmetry breaking.
Long-range contributions to $\znbb$ decay arise in such models
\cite{Hirsch:1996ye}.

Neutrinoless double beta decay may impose constraints on all
(non-gauge) vectors in table \ref{tab:Groups_for_each_rep} provided
the corresponding models violate lepton number. We will not discuss
here possible models nor give individual limits case by case. However,
we do want to briefly discuss the typical limits one expects for any
such model. For this estimate, one has to distinguish models leading
to a contribution to the long-range $\znbb$ decay amplitude from those
that contribute only to the short-range part.

The following subset of vectors can contribute to the long-range
amplitude at the level of $d=7$ operators: $\left(\mathbf{3},\mathbf{1},2/3\right)$,
$\left(\mathbf{3},\mathbf{3},2/3\right)$,
$\left(\mathbf{3},\mathbf{2},1/6\right)$, $\left(\mathbf{1},\mathbf{2},3/2\right)$
and $\left(\mathbf{1},\mathbf{1},1\right)$.
Constraints on $d=7$ operators are always of the form
${\cal O}_{d=7}^{long} \propto \frac{g_{eff}^3 v}{\Lambda_{LNV}^3}$,
where $g_{eff}$ is the geometric mean of the three couplings entering
the decomposition of the relevant operator, and $\Lambda_{LNV}$ is the
geometric mean of the masses of the heavy particles. Using the latest
experimental half-life for $^{136}$Xe \cite{KamLAND-Zen:2016pfg} and
including QCD corrections \cite{Arbelaez:2016zlt} in the calculation
of the theoretical amplitude, lower limits in the range of
$\Lambda_{LNV} \gsim g_{eff} (26-247)$ TeV are found, depending on
which of the coefficients in the general expression for the $\znbb$
decay half-life is generated in the corresponding model. We want to
stress that in some particular cases one can expect $g_{eff} \ll 1$,
leading to much more relaxed bounds on $\Lambda_{LNV}$. A well-known
example is the left-right symmetric model (see section
\ref{subsect:LR}).  Many more vectors appear in the long-range part of
$\znbb$ if we include operators up to $d=9$. However, constraints are
much weaker for these operators: ${\cal O}_{d=9}^{long} \propto
\frac{g_{eff}^5 v^3}{\Lambda_{LNV}^5}$.  This leads to bounds of the
order of $\Lambda_{LNV}\gsim (3.5-13.5) g_{eff}$ TeV, again depending
on the operator. We will discuss some examples of long-range ${\cal
	O}_{d=9}$ contribution (from gauged vectors) to $\znbb$ decay in
sections \ref{subsect:LR} and \ref{subsect:331}.

In order to generate short-range contributions from exotic vectors a
full model requires either an additional exotic scalar (topology II),
or an exotic fermion (topology I). We will discuss a gauged example
for topology II in section \ref{subsect:331} and some example diagrams
for topology I in \ref{subsect:LR} and \ref{subsect:422}. Contributions 
to $\znbb$ decay via $d=9$ operators scale as 
${\cal O}_{d=9}^{short}\propto\frac{g_{eff}^{4}}{\Lambda_{LNV}^{5}}$,
assuming that the triple vector vertex has a coupling of the order
$g_{eff} \Lambda_{LNV}$. Taking the general short-range decay rate,
including QCD corrections, from
\cite{Gonzalez:2015ady,Arbelaez:2016uto} and using the limit from
$^{136}$Xe \cite{KamLAND-Zen:2016pfg}, results in lower limits in the
range $\Lambda_{LNV}\gsim (2.2-7.3) g_{eff}^{4/5}$ TeV, depending on
the operator.

These generic limits relating the geometric mean mass $\Lambda_{LNV}$
and the geometric mean coupling $g_{eff}$ are to be understood 
only as rough order of magnitude estimates. For specific models and diagrams, 
the masses and couplings themselves can have different values spanning
several orders of magnitude, as we will discuss in the 
following sections. However, all numbers given in the examples below are consistent 
with the above estimates.

\subsection{A well-known case: Left-right symmetric model}
\label{subsect:LR}

The vector representation $\left(\mathbf{1},\mathbf{1},1\right)$ can
be identified with the charged component of a right triplet
$\left(\mathbf{1},\mathbf{1},\mathbf{3},0\right)$ of the left-right
symmetric group $SU(3)_c\times SU(2)_L\times SU(2)_R\times U(1)_{B-L}$.
Neutrinoless double beta decay was discussed
in this context first in \cite{Mohapatra:1980yp}. Many more detailed
calculations, with particular emphasis on the nuclear physics side of
the problem, followed soon thereafter (see
\cite{Haxton:1985am,Doi:1985dx} for reviews).  These early
calculations focused on long-range contributions, but the importance
of the short-range part was stressed again in \cite{Mohapatra:1986pj}.
The general decay rate in left-right models, including long-range and
short-range diagrams, was then given in
\cite{Doi:1992dm,Hirsch:1996qw}. The latter reference not only
calculated nuclear matrix elements but also included for the first
time diagrams with a scalar $\Delta_R \equiv
\left(\mathbf{1},\mathbf{1},\mathbf{3},1\right)$ \cite{Hirsch:1996qw}.
A more recent review on $\znbb$ decay in left-right symmetric models
can be found in \cite{Rodejohann:2011mu}. Our discussion below of $\znbb$
decay in left-right models follows essentially
\cite{Helo:2016vsi}.

In the standard left-right symmetric model, the $SU(2)_R\times U(1)_{B-L}$ group is
broken to $U(1)_Y$ via the vacuum expectation value (vev) of the
scalar $\Delta_R$, and Majorana neutrino masses for the right-handed
neutrinos are generated from the same vev.  Left-handed neutrino
masses are then produced after electro-weak symmetry breaking through
a seesaw mechanism.  As noted above, in this setup both long-range and
short-range contributions to $\znbb$ decay have to be considered ---
see the diagrams in figure \ref{fig:LRdiags}.

\begin{center}
	\begin{figure}[tbph]
		\begin{centering}
			\includegraphics[scale=0.7]{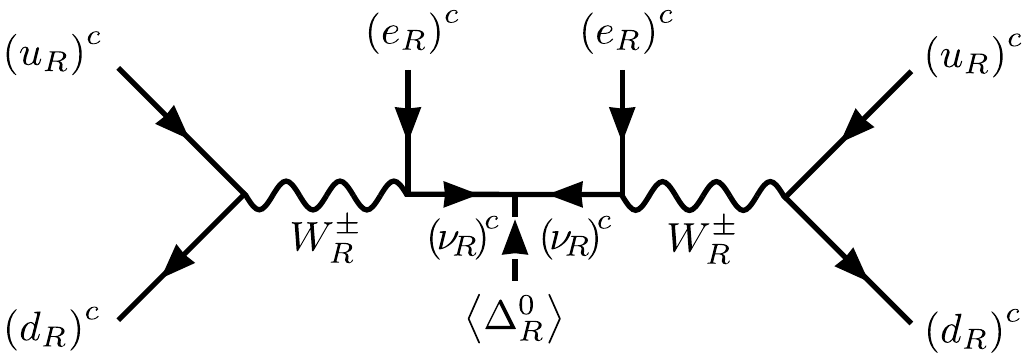} \\ \vskip5mm
			\hskip-5mm\includegraphics[scale=0.7]{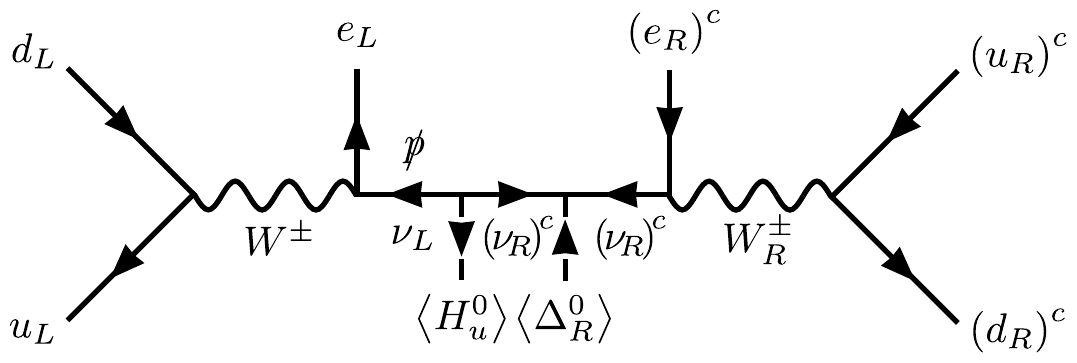} \hskip5mm
			\includegraphics[scale=0.7]{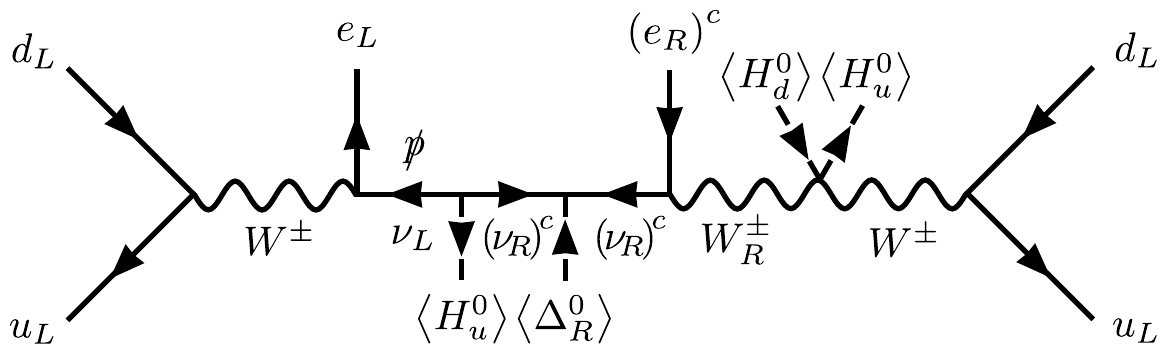}
			\par
		\end{centering}
		\protect\caption{\label{fig:LRdiags} $\znbb$ decay diagrams in the
			left-right symmetric model. Top: Short-range contribution due to
			heavy right-handed neutrino exchange. Bottom: Long-range
			contributions due to left-right neutrino mixing.}
	\end{figure}
	\par
\end{center}

\noindent
The  $d=7$ and $d=9$ operators associated to these diagrams are the following:
\begin{gather}
{\cal O}_{8}=\overline{L}e^{c}u^{c}\overline{d^{c}}H\,,\,{\cal O}_{7}=\overline{L}e^{c}Q\overline{Q}H^*H^*H^*\,,\\
{\cal O}_{-}=e^{c}e^{c}u^{c}u^{c}\overline{d^{c}}\overline{d^{c}}.
\end{gather}
Consider first the short-range diagram generated by ${\cal O}_-$.
This diagram is proportional to $\frac{g_{R}^4}{m_{W_R}^4\langle
	m_N\rangle}$, where $\langle m_N\rangle$ is defined as $1/\langle
m_N\rangle = \sum^6_{j=4} V_{ej}^2/m_j$ and in turn $V_{ej}$ is the right-handed
neutrino mixing matrix in the notation of \cite{Doi:1985dx} and the
index $j=4,5,6$ runs over the heavy neutrino mass eigenstates. Similar
to the case of the well-known mass mechanism, there can be
cancellations among terms in the sum over $j$, due to Majorana phases
in the right-handed neutrino sector. However, $\sum^6_{j=4} V_{ej}^2 \simeq
1$ from unitarity. Updating \cite{Hirsch:1996qw} with the latest 
half-life bound from \cite{KamLAND-Zen:2016pfg} results in a lower 
limit on the $W_R$ boson mass, roughly given by $m_{W_R}\gsim 
1.9 (g_R/g_L) (\frac{1 \hskip1mm \rm TeV}{\langle m_N\rangle})$ TeV.

For the long-range part of the amplitude, one finds two different
contributions, which are shown in figure \ref{fig:LRdiags}.  One can identify these
diagrams with the $\lambda$ and $\eta$ terms in the notation of
\cite{Doi:1985dx}:
\begin{eqnarray}\label{eq:efflam}
{\cal O}_8 & \to & \langle\lambda\rangle = \sum_{j=1}^3 U_{ej}V_{ej}\lambda 
= \sum_{j=1}^3 U_{ej}V_{ej} \left(\frac{m_{W_L}}{m_{W_R}}\right)^2 ,
\\ \label{eq:effeta}
{\cal O}_7 & \to & \langle\eta\rangle = \sum_{j=1}^3 U_{ej}V_{ej}\eta 
= \sum_{j=1}^3 U_{ej}V_{ej}\tan\zeta .
\end{eqnarray}
Here, $\zeta$ is the mixing angle between the $W_L$ and $W_R$ states,
typically also roughly of the order of $\left(\frac{m_{W_L}}{m_{W_R}}\right)^2$.
$U_{\alpha j}$ is the left-handed neutrino mixing matrix and
orthogonality of $U_{ej}$ and $V_{ej}$ implies $\sum_{j=1}^6
U_{ej}V_{ej} \equiv 0$. However, the sums in equations (\ref{eq:efflam}) and
(\ref{eq:effeta}) run only over the light neutrinos and thus, are
incomplete. Typical expectations for this heavy-light neutrino
mixing, $\sum_{j=1}^3 U_{ej}V_{ej}$, in models with an ordinary type-I
seesaw are of the order of $m_D/M_M \sim \sqrt{m_{\nu}/M_M}$, where
$m_D$, $M_M$ and $m_{\nu}$ are the Dirac mass term, the Majorana mass
term for the right-handed neutrinos and the light neutrino mass,
respectively. Using again the limit from $^{136}$Xe 
\cite{KamLAND-Zen:2016pfg} results in upper limits for 
$\langle\lambda\rangle \lsim 2.1 \times 10^{-7}$ and 
$\langle\eta\rangle \lsim 1.1 \times 10^{-9}$ using the 
nuclear matrix elements of either \cite{Muto:1989cd} or \cite{Stefanik:2015twa}.\footnote{
	It is amusing to note that using the nuclear matrix elements of either
	\cite{Muto:1989cd} or \cite{Stefanik:2015twa} results in nearly
	identical numerical limits.}  For $\sum_{j=1}^3 U_{ej}V_{ej} \gsim
10^{-4}$ the limit on $\langle\lambda\rangle$ would become a more
stringent limit on $m_{W_R}$ than the one derived from the short range
diagram. However, for $\langle m_N\rangle \simeq 1$ TeV one would
expect $\sum_{j=1}^3 U_{ej}V_{ej} \sim 10^{-6}$.  It is interesting to
note \cite{Helo:2016vsi}, that the constraint on ${\cal O}_7$ is
actually more stringent than the one on ${\cal O}_8$, despite being
formally of higher order, due to the contribution of the nuclear
recoil matrix element to the former \cite{Doi:1985dx}.

\subsection{Models based on the 331 group}
\label{subsect:331}

In this section, we discuss examples of exotic contributions based on
the $SU(3)_C\times SU(3)_L \times U(1)_X$ group. It turns out that there is a rich variety of such models, and the most important distinguishing factor among them is the way the SM hypercharge is embedded
in $SU(3)_L \times U(1)_X$: $Y=\beta T_8 + X$.  From purely group
theoretical considerations, $\beta$ is a continuous parameter.  However, perturbativity of the gauge interactions requires that $\beta \lsim 1.8$ 
(see for example \cite{Dias:2004dc,Fonseca:2016xsy}). Furthermore, if one additionally demands that there are no
fractionally charged color singlet fermions in the model, $\beta$ may
take only the values $\pm 1/\sqrt{3},\pm\sqrt{3}$.

In 331 models, usually all three generations of left-handed leptons are
assigned to triplets. Anomaly cancellation then requires that two
left-handed quark generations are assigned to anti-triplets, while the third one must be a
triplet.\footnote{Alternatively one can put all left-handed quark generations into
	anti-triplets.  In this ``flipped'' scenario, two left-handed lepton generations are
	assigned to triplets, while the remaining one is part of an $SU(3)_L$
	sextet \cite{Fonseca:2016tbn}.}  The prototype 331 model for this
kind of construction (with $\beta=-1/\sqrt{3}$) was proposed in
\cite{Singer:1980sw} and \cite{Valle:1983dk}. The only difference
between the variants \cite{Singer:1980sw} and \cite{Valle:1983dk} is
that the former assumes the presence of additional right-handed neutral singlets
$\nu_R$, while the
latter explicitly excludes such states from the fermion spectrum.

For the values $\beta=\pm 1/\sqrt{3}$ and $\beta=\pm\sqrt{3}$, some of the gauge bosons transform
as $\left(\mathbf{1},\mathbf{2},\frac{1}{2}\right)$ and $\left(\mathbf{1},\mathbf{2},\frac{3}{2}\right)$, respectively,
under the SM gauge group, and both these representations were identified earlier as potentially important for neutrinoless
double beta decay. However, even though it is possible to build one or more models for each of these values of $\beta$,
from the point of view of $\znbb$ we find that $\beta=-\sqrt{3}$ is the most interesting choice, given that the third element
of the lepton $SU(3)_L$ triplets corresponds to the SM right-handed charged leptons $\ell^c_{\alpha}$. The significance of this
fact will be discussed shortly.

We will then concentrate on the 331 model based on the choice
$\beta=-\sqrt{3}$ and the fields shown in table \ref{tab:PP-reps}. This variant was first proposed in
\cite{Pisano:1991ee,Frampton:1992wt}, and therefore we will call it the
Pisano-Pleitez-Frampton (PPF) model.

\begin{table}[tbph]
	\begin{centering}
		\begin{tabular}{cccccc}
			\toprule 
			Field  & \# flavours  & 331 representation  & $G_{SM}$ decomposition  & Components  & Lepton number\tabularnewline
			\midrule 
			$\psi_{\ell,\alpha}$  & 3  & $\left(\mathbf{1},\mathbf{3},0\right)$  & $\left(\mathbf{1},\widehat{\mathbf{2}},-\frac{1}{2}\right)+\left(\mathbf{1},\widehat{\mathbf{1}},1\right)$  & $\left(\left(\nu_{\alpha},\ell_{\alpha}\right),\ell_{\alpha}^{c}\right)^{T}$  & $\left(1,1,-1\right)^{T}$\tabularnewline
			$Q_{\alpha=1,2}$  & 2  & $\left(\mathbf{3},\overline{\mathbf{3}},-\frac{1}{3}\right)$  & $\left(\mathbf{3},\widehat{\mathbf{2}},\frac{1}{6}\right)+\left(\mathbf{3},\widehat{\mathbf{1}},-\frac{4}{3}\right)$  & $\left(\left(d_{\alpha},-u_{\alpha}\right),J_{\alpha}^{c}\right)^{T}$  & $\left(0,0,2\right)^{T}$\tabularnewline
			$Q_{3}$  & 1  & $\left(\mathbf{3},\mathbf{3},\frac{2}{3}\right)$  & $\left(\mathbf{3},\widehat{\mathbf{2}},\frac{1}{6}\right)+\left(\mathbf{3},\widehat{\mathbf{1}},\frac{5}{3}\right)$  & $\left(\left(t,b\right),J_{3}^{c}\right)^{T}$  & $\left(0,0,-2\right)^{T}$\tabularnewline
			$u_{\alpha}^{c}$  & 3  & $\left(\overline{\mathbf{3}},\mathbf{1},-\frac{2}{3}\right)$  & $\left(\overline{\mathbf{3}},\widehat{\mathbf{1}},-\frac{2}{3}\right)$  & $u_{\alpha}^{c}$  & 0\tabularnewline
			$d_{\alpha}^{c}$  & 3  & $\left(\overline{\mathbf{3}},\mathbf{1},\frac{1}{3}\right)$  & $\left(\overline{\mathbf{3}},\widehat{\mathbf{1}},\frac{1}{3}\right)$  & $d_{\alpha}^{c}$  & 0\tabularnewline
			$J_{\alpha=1,2}$  & 2  & $\left(\overline{\mathbf{3}},\mathbf{1},\frac{4}{3}\right)$  & $\left(\overline{\mathbf{3}},\widehat{\mathbf{1}},\frac{4}{3}\right)$  & $J_{\alpha}$  & -2\tabularnewline
			$J_{3}$  & 1  & $\left(\overline{\mathbf{3}},\mathbf{1},-\frac{5}{3}\right)$  & $\left(\overline{\mathbf{3}},\widehat{\mathbf{1}},-\frac{5}{3}\right)$  & $J_{3}$  & 2\tabularnewline
			$\phi_{1}$  & 1  & $\left(\mathbf{1},\mathbf{3},1\right)$  & $\left(\mathbf{1},\widehat{\mathbf{2}},\frac{1}{2}\right)+\left(\mathbf{1},\widehat{\mathbf{1}},2\right)$  & $\left(\left(\phi_{1}^{+},\phi_{1}^{0}\right),\widetilde{\phi}_{1}^{++}\right)^{T}$  & $\left(0,0,-2\right)^{T}$\tabularnewline
			$\phi_{2}$  & 1  & $\left(\mathbf{1},\mathbf{3},-1\right)$  & $\left(\mathbf{1},\widehat{\mathbf{2}},-\frac{3}{2}\right)+\left(\mathbf{1},\widehat{\mathbf{1}},0\right)$  & $\left(\left(\phi_{2}^{-},\phi_{2}^{--}\right),\widetilde{\phi}_{2}^{0}\right)^{T}$  & $\left(2,2,0\right)^{T}$\tabularnewline
			$\phi_{3}$  & 1  & $\left(\mathbf{1},\mathbf{3},0\right)$  & $\left(\mathbf{1},\widehat{\mathbf{2}},-\frac{1}{2}\right)+\left(\mathbf{1},\widehat{\mathbf{1}},1\right)$  & $\left(\left(\phi_{3}^{0},\phi_{3}^{-}\right),\widetilde{\phi}_{3}^{+}\right)^{T}$  & $\left(0,0,-2\right)^{T}$\tabularnewline
			\bottomrule
		\end{tabular}
		\par\end{centering}
	
	\protect
	\caption{\label{tab:PP-reps}Fields in the Pisano-Pleitez-Frampton 
		(PPF) model \cite{Pisano:1991ee,Frampton:1992wt}, which is based on the  $SU(3)_C\times SU(3)_L \times U(1)_X$ gauge group.}
\end{table}

As shown in table \ref{tab:PP-reps}, the standard model lepton doublets are in
$SU(3)_L$ triplets, $\psi_{\ell,\alpha}$, with $\alpha=e,\mu,\tau$, which also contain a
singlet charged lepton, $\ell^c_{\alpha}$. Note that our conventions are
that all fields are left-handed, meaning that $\ell_\alpha^c$ corresponds to the
charged-conjugate of the right-handed leptons. It is of utmost importance to correctly identify the source or sources of lepton number violation if we are to discuss
neutrinoless double beta decay and, as explained in
detail in \cite{Fonseca:2016xsy}, the original PPF model contains a single lepton number violating interaction:
\begin{equation}\label{eq:LNVPPF}
{\mathscr{L}}^{LNV} = \lambda_7 \phi_1\phi_2\phi_3^*\phi_3^* + \textrm{h.c.}\,.
\end{equation}

The original PPF model predicts that the charged lepton masses are
($-m,0,m$), in gross violation of the experimental data, and thus has
to be modified. There are two simple possibilities to
generate a realistic charged lepton spectrum. One of them
is to introduce a vector-like pair of singlet charged fermions
\cite{Duong:1993zn,Montero:2001ji}.  We will call this version of the model
PPF-E. The other possibility involves adding a scalar $SU(3)_L$ sextet with no $U(1)_X$ charge. This
variant was first proposed in \cite{Foot:1992rh}, and we will call it
PPF-S.

Consider first the addition of a vector-like lepton singlet $E$/$E^c$, which is allowed to have the following interactions:
\begin{equation}\label{eq:LagEc}
\mathscr{L}_{EE^c} = h_{E^c,\alpha} \psi_{\ell,\alpha} E^c \phi_1^* 
+ h_{E,\alpha} \psi_{\ell,\alpha} {E} \phi_2^* + m_{EE^c} E{E^c}\,.
\end{equation}
Once $\phi_{1,2}$ obtain vevs, the SM charged leptons mix with the 
vector-like state and the charged lepton masses can be made to agree with the 
experimental data. As discussed in \cite{Fonseca:2016xsy}, correctly fitting 
the $\tau$ mass requires $|h_{E^c,\tau} h_{{E},\tau}| \sim {\cal O}(10^{-2})$ 
for $m_{EE^c}\sim $ TeV. 

The PPF-E model also makes it possible to fit neutrino data, since the interactions shown in equations (\ref{eq:LNVPPF}) and (\ref{eq:LagEc}) can be used to generate the
1-loop neutrino mass diagram shown in figure \ref{fig:numass}.  A
simple estimate of this loop gives \cite{Fonseca:2016xsy}
\begin{equation}\label{eq:diagC}
(m_{\nu})_{\alpha\beta} 
\sim \left(\frac{m_{E E^c}}{\rm TeV} \right)  
\left(\frac{\rm TeV}{{m_{\phi_{1,2}}}}\right)^2
\left(\frac{\left\langle \phi_{3}^{0}\right\rangle}{100 \hskip1mm {\rm GeV}}\right)^2
\left(\frac{
	(|h_{E^c,\alpha} h_{{E},\beta} + h_{E^c,\beta} h_{{E},\alpha}|}{10^{-2}}
\right)
\left(\frac{\lambda_7}{10^{-7}}\right) 10^{-1}\hskip 1mm {\rm eV}\,.
\end{equation}
Note that since $\meff = \sum_j U_{ej}^2m_j \equiv (m_{\nu})_{ee}$, neutrinoless double beta decay is sensitive only to
$(m_{\nu})_{ee}$, \textit{i.e.} to the entry of the neutrino mass where $\alpha=\beta=e$.

\begin{center}
	\begin{figure}[tbph]
		\begin{centering}
			\includegraphics[scale=0.8]{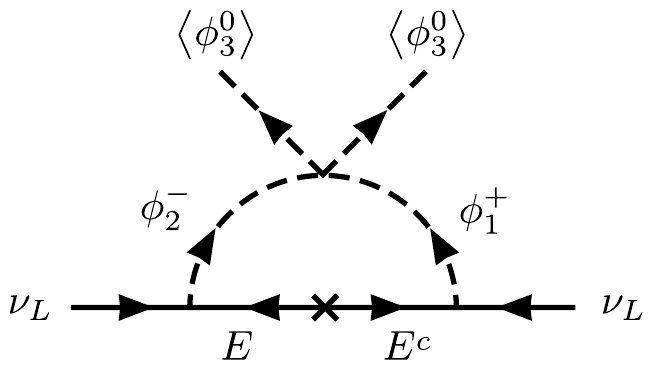}
			\par
		\end{centering}
		\protect\caption{\label{fig:numass} One loop
			neutrino mass diagram in the PPF-E model.}
	\end{figure}
	\par
\end{center}

\noindent
The second variation of PPF model mentioned above involves adding a scalar
sextet, $S=\left(\mathbf{1},\mathbf{6},0\right)$:
\begin{equation}\label{eq:sexplet}
S = 
\begin{pmatrix}
\Delta^{0} & \frac{1}{\sqrt{2}}\Delta^{-} & \frac{1}{\sqrt{2}}H^{+}_S \\
\frac{1}{\sqrt{2}}\Delta^{-} & \Delta^{--} & \frac{1}{\sqrt{2}}H^{0}_S \\
\frac{1}{\sqrt{2}}H^{+}_S & \frac{1}{\sqrt{2}}H^{0}_S & \sigma^{++}
\end{pmatrix}\,.
\end{equation}
$S$ decomposes into a triplet, a doublet, and a singlet of $SU(2)_L$, hence we have named the corresponding components with the
names  $\Delta$, $H_S$, and $\sigma$, respectively. The interaction term of the sextet with the leptons decomposes as follows:
\begin{equation}\label{eq:sxpint}
y_S \psi_{\ell,\alpha} S^* \psi_{\ell,\beta} 
= y_S \left[ \nu_{\alpha}\nu_{\beta}\Delta^{0*}
+ \frac{1}{\sqrt{2}} 
(\ell_{\alpha}\ell^{c}_\beta+\ell^{c}_\alpha \ell_{\beta})H^{0*} 
+ \cdots \right]\,,
\end{equation}
and from this expression it is obvious that a non-zero $\Delta^0$ vev will give rise to a type-II seesaw. This vev necessarily violates lepton number by two units ($\Delta L=2$).\footnote{Note that it is also possible to write down new scalar interactions involving $S$ which break explicitly lepton number by two units.}  The charged leptons now receive two
contributions to their mass: $m_\ell = y_\ell \langle \phi^0_3\rangle
+ y_S \langle H^0_S\rangle$. To generate a large enough $m_{\tau}$,
this requires that either $(y_S)_{\mu\tau} \langle H^0_S\rangle$
and/or $(y_S)_{\tau\tau} \langle H^0_S\rangle$ are of the order of
$m_{\tau}$, which puts a lower limit of roughly ${\cal O}(10^{-2})$ on
at least one of these Yukawa couplings.

This is important since the same Yukawa couplings contribute to the Majorana
neutrino mass matrix and thus one must have
$\langle\Delta^0\rangle/\langle H_S^0\rangle \lsim 10^{-10}$ in order
not to violate upper bounds on the neutrino mass scale. Given this
seemingly strong constraint, one might wonder whether it is possible
to find a symmetry that produces an exactly vanishing triplet vev, $\langle\Delta^0\rangle=0$,
as was done in the original work \cite{Foot:1992rh}. The answer is that such symmetry would also eliminate all lepton
number violating terms from the PPF-S model \cite{Fonseca:2016xsy}.
This is thus not an acceptable solution, since neutrinos would be
massless.   On the other hand, it is possible to
arrange a small but non-zero ratio of vevs $\langle\Delta^0\rangle/\langle H_S^0\rangle$ with a proper choice of scalar potential parameters; some of them would have to be small though (for example the coefficient of the trilinear interaction $\phi_3 \phi_3 S^*$). In any case, we will assume a small $\langle\Delta^0\rangle\ne 0$ in the following discussion.

\begin{center}
	\begin{figure}[tbph]
		\begin{centering}
			\includegraphics[scale=0.8]{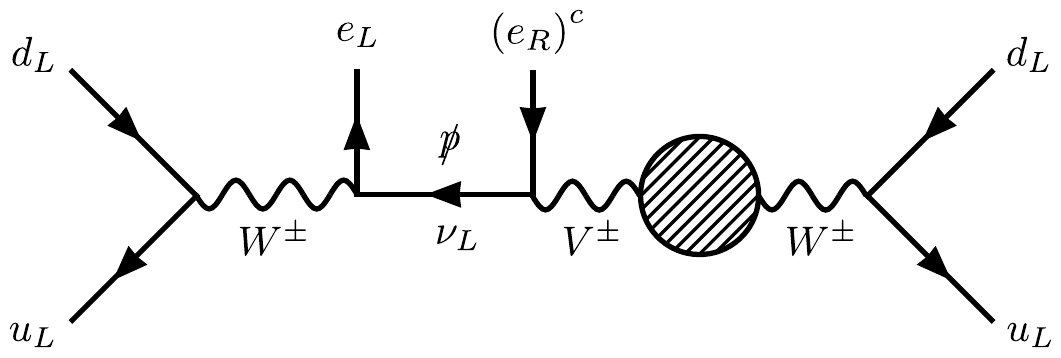}
			\par
		\end{centering}
		\protect\caption{\label{fig:LR331}
			Long-range contribution to $\znbb$ decay in extended PPF 331 models.}
	\end{figure}
	\par
\end{center}

In addition to the mass mechanism, the PPF model and its variants may produce other
long-range contributions to $\znbb$ decay, which are associated to the diagram shown in figure \ref{fig:LR331}. To understand the physics behind this diagram,
we first have to discuss charged-current interactions.  Apart from the
ordinary $W^{\pm}$ boson which couples to left-handed doublets, the PPF model
contains exotic heavy charged bosons which, following
\cite{Pisano:1991ee}, we will denote as $V^{\pm}$ and $U^{\pm\pm}$.
The charged current interactions then contain the terms:
\begin{align}
\mathscr{L}_{L}^{cc}= & -\frac{g_{L}}{\sqrt{2}}\sum_{\alpha=1}^{3}\left(\overline{\nu}_{\alpha}\gamma^{\mu}\ell_{\alpha}W_{\mu}^{+}+\overline{\ell}_{\alpha}^{c}\gamma^{\mu}\nu_{\alpha}V_{\mu}^{+}+\overline{\ell}_{\alpha}^{c}\gamma^{\mu}\ell_{\alpha}U_{\mu}^{++}+\textrm{h.c.}\right)\,,\label{eq:ccppf1}\\
\mathscr{L}_{Q_{\alpha=1,2}}^{cc}= & -\frac{g_{L}}{\sqrt{2}}\sum_{\alpha=1}^{2}\Big(\overline{u}_{\alpha}\gamma^{\mu}d_{\alpha}W_{\mu}^{+}-\overline{d}_{\alpha}\gamma^{\mu}J_{\alpha}^{c}V_{\mu}^{+}+\overline{u}_{\alpha}\gamma^{\mu}J_{\alpha}^{c}U_{\mu}^{++}+\textrm{h.c.}\Big)\,,\label{eq:ccppf2}\\
\mathscr{L}_{Q_{3}}^{cc}= & -\frac{g_{L}}{\sqrt{2}}\Big(\overline{t}\gamma^{\mu}bW_{\mu}^{+}+\overline{J}_{3}^{c}\gamma^{\mu}tV_{\mu}^{+}+\overline{J}_{3}^{c}\gamma^{\mu}bU_{\mu}^{++}+\textrm{h.c.}\Big)\,.\label{eq:ccppf3}
\end{align}
It is important that $V^{\pm}$ couples only ``off-diagonally'' to
quarks, \textit{i.e.} it always connects one SM quark with one of the exotic
states $J^c_i$. On the other hand, as a consequence of equations (\ref{eq:ccppf1})--(\ref{eq:ccppf3}), the
neutrino propagator between a $W^{+}$ and a $V^{+}$ vertex is
proportional to
\begin{equation}\label{eq:prop}
P_{L}\left(m_{\nu_{i}}+\pslash\right)P_{R}=\pslash P_{R}\,.
\end{equation}
Recall that the magnitude of the $\pslash$ matrix is associated to the Fermi momentum of the nucleons, $(100-200)$ MeV, which is ${\cal
	O}(10^9)$ times larger than the average neutrino mass. However, in order to complete the diagram in
figure \ref{fig:LR331}, a $\Delta L=2$ interaction is needed. (The
$\pslash$ part of the (Majorana) neutrino propagator does not violate
lepton number.) Similarly to the situation encountered in the left-right
symmetric model, such a $\Delta L=2$ term can be generated by
$V^\pm-W^\pm$ mixing. As shown in figure \ref{fig:LRball}, there are 
different possibilities to generate gauge boson mixing in the PPF 331. 
We will briefly discuss each of these possibilities in turn.

\begin{center}
	\begin{figure}[tbph]
		\begin{centering}
			\includegraphics[scale=0.8]{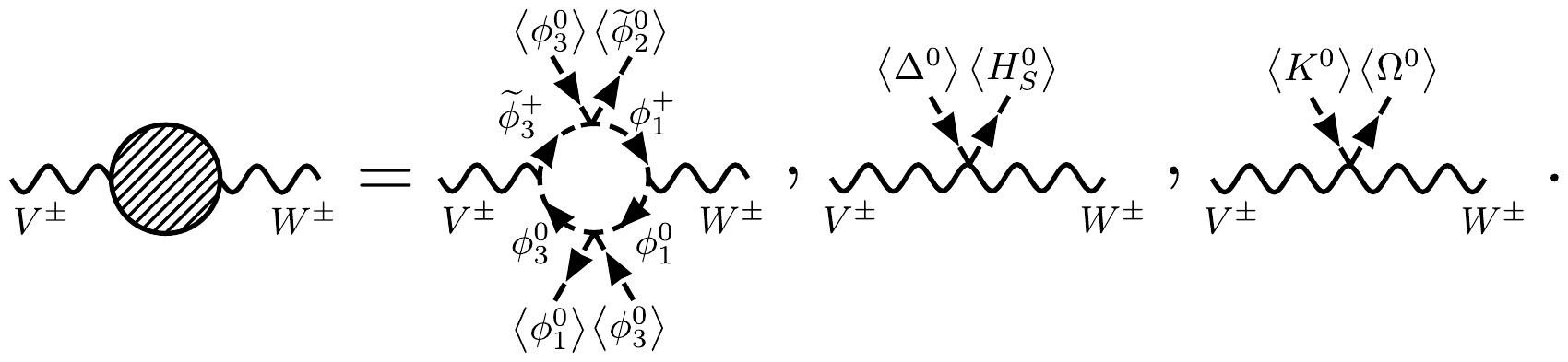}
			\par
		\end{centering}
		\protect\caption{\label{fig:LRball}
			Different realizations of gauge boson mixing in PPF 331 models. 
			From left to right: PPF-E, PPF-S, and a PPF variant with a scalar field 
			transforming as a $\boldsymbol{10}$ under $SU(3)_L$. This decuplet decomposes into
			a quadruplet $K$, a triplet $\Omega$, and other representations of $SU(2)_L$.}
	\end{figure}
	\par
\end{center}

In the PPF-E model, $V^\pm-W^\pm$ mixing is generated through a loop which is shown
in figure \ref{fig:LR331}. On top of the quartic interaction in equation (\ref{eq:LNVPPF}), this loop requires one of the
following terms in the scalar potential:
\begin{eqnarray}\label{eq:spot}
{\mathscr{L}}_S = \mu \phi_1\phi_2\phi_3 + \lambda_4 |\phi_1|^2|\phi_2|^2
+ \lambda_5  |\phi_1|^2|\phi_3|^2
+ \lambda_6  |\phi_2|^2|\phi_3|^2 + \cdots\,.
\end{eqnarray}
Note that while only one possible diagram contributing to $V^+-W^+$ mixing
is shown in the figure, permutations of this loop using different
$\lambda_k$, $k=4,5,6$ or two terms proportional to $\mu$ all
contribute. One can make a rough estimate of the size of the 
mixing angle:
\begin{equation}\label{eq:mix}
\theta_{VW} \sim \frac{g_L^2}{16\pi^2}\lambda_7 \lambda_{k}\left(\frac{v}{n}\right)^{3}
\end{equation}
Here, $k=4,5,6$ and we assume that all the doublet vevs $\langle
\phi^0_{1,3}\rangle$ are of the order of the
SM vev, $v \simeq 174$ GeV, while the SM singlet vev $\langle\widetilde{\phi}_{2}^{0}\rangle$ is of the order of the 331 scale, denoted as $n$. For
$n$ of the order of one TeV, and using the upper limit of $C_3^L=1.9
\times 10^{-9}$ \cite{Arbelaez:2016zlt}, one gets
a bound $\lambda_7\lambda_k \lsim 10^{-4}$ which is to be compared 
with the limit on $\lambda_7$ obtained from the absolute neutrino mass scale --- see
equation  (\ref{eq:diagC}). This later constraint seems more important, even 
though it depends on extra parameters such as $h_E$ and $h_{E^c}$. 

For the PPF-S model (see figure \ref{fig:LRball}) the estimate of the mixing angle is simply $\theta_{VW} \sim
g_L^2 \langle \Delta^0\rangle \langle H^0_S\rangle/n^2$. As discussed above, the triplet vev must be tiny due to the upper limit on the neutrino masses and
the simultaneous need to obtain the correct charged lepton masses. Thus, from a non-zero $\left\langle \Delta^{0}\right\rangle 
$ we get $\theta_{VW} \lsim 10^{-12}$ in the PPF-S model and long-range
contributions to $\znbb$ decay due to this vev are completely negligible, contrary 
to the claims made in \cite{Pleitez:1993gc,Montero:2000ar}.

Finally, we would like to briefly comment that $V^\pm-W^\pm$ mixing can easily be
made much larger in non-minimal models. For example, let us
consider adding a ${\bf 10}$-plet of $SU(3)_L$ to the PPF model. Under the SM
group this field decomposes as:
\begin{equation}
(\mathbf{1},\mathbf{10},0)\to\left(\mathbf{1},\mathbf{4},-\frac{3}{2}\right)+(\mathbf{1},\mathbf{3},0)+\left(\mathbf{1},\mathbf{2},\frac{3}{2}\right)+(\mathbf{1},\mathbf{1},3)=K+\Omega+\cdots\,.
\end{equation}
Both $K$ and the $\Omega$ contain neutral components, which 
may acquire vevs: $\langle K^0\rangle$ and $\langle \Omega^0\rangle$. 
Non-zero values for $\langle K^0\rangle$ and $\langle \Omega^0\rangle$ 
generate $V^\pm-W^\pm$ mixing, as indicated in figure  \ref{fig:LRball}. 

Before estimating the constraints on $V^\pm-W^\pm$ mixing, let us briefly
comment on the $\rho$ parameter. Including the ${\bf 10}$-plet, in the
limit $n \gg v_{SM}$, the $W^{\pm}$ and $Z^0$ masses are given by
\begin{align}
m_{W^{\pm}}^{2} & =\frac{1}{2}g_{L}^{2}\left(\left\langle \phi_{1}^{0}\right\rangle ^{2}+\left\langle \phi_{3}^{0}\right\rangle ^{2}+3\left\langle K^{0}\right\rangle ^{2}+4\left\langle \Omega^{0}\right\rangle ^{2}\right)\,,\label{eq:mass10_1}\\
m_{Z^{0}}^{2} & =\frac{1}{2}\left(g_{L}^{2}+g_{Y}^{2}\right)\left(\left\langle \phi_{1}^{0}\right\rangle ^{2}+\left\langle \phi_{3}^{0}\right\rangle ^{2}+9\langle K^{0}\rangle^{2}\right)\,.\label{eq:mass10_2}
\end{align}
Taking the value $\rho = 1.00040 \pm 0.00024$ from
\cite{Beringer:1900zz}, these formulas imply that one has the $2\sigma$ limits  $\langle
K^0\rangle \le 1.3$ GeV for $\langle \Omega^0\rangle \equiv 0$, and
$\langle \Omega^0\rangle \le 2.9$ GeV for $\langle K^0\rangle \equiv
0$. Note that there is a special direction, $\langle\Omega^0\rangle^2 =3/2
\langle K^0\rangle^2$, for which no constraint on the vevs can be
derived from $\rho$. 

Absence of $\znbb$ decay puts an upper bound on $\theta_{VW}$ also in this
setup. From $C_3^L=1.9 \times 10^{-9}$ \cite{Arbelaez:2016zlt} one can
estimate $\langle \Omega^0\rangle \langle K^0\rangle \lsim 0.0019 $
GeV$^2$ for $n=1$ TeV. This constraint is much more stringent than 
the one derived from the $\rho$ parameter.

\begin{center}
	\begin{figure}[tbph]
		\begin{centering}
			\includegraphics[scale=0.8]{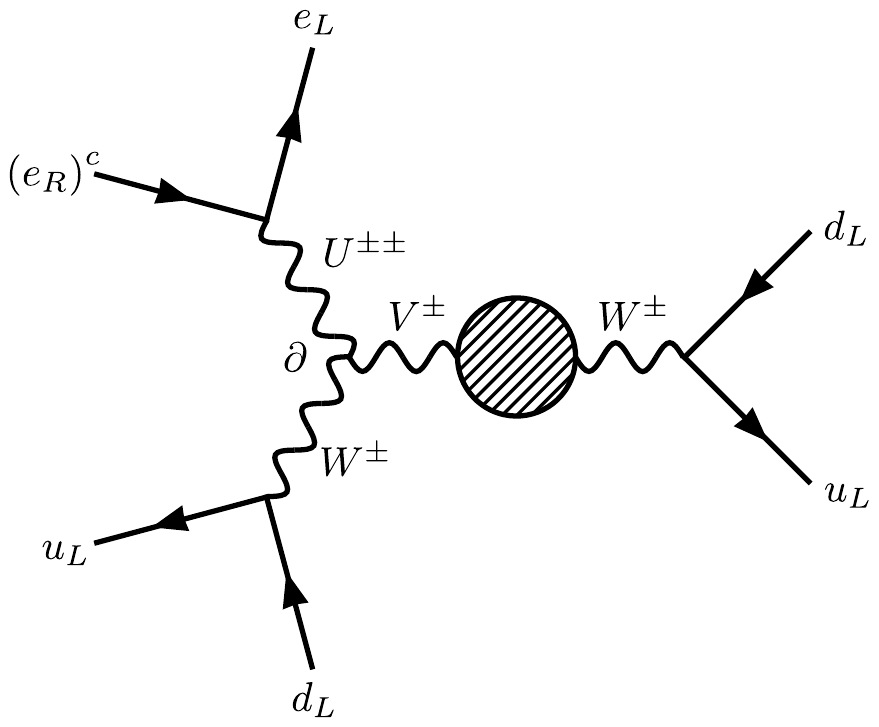}\hskip10mm
			\includegraphics[scale=0.8]{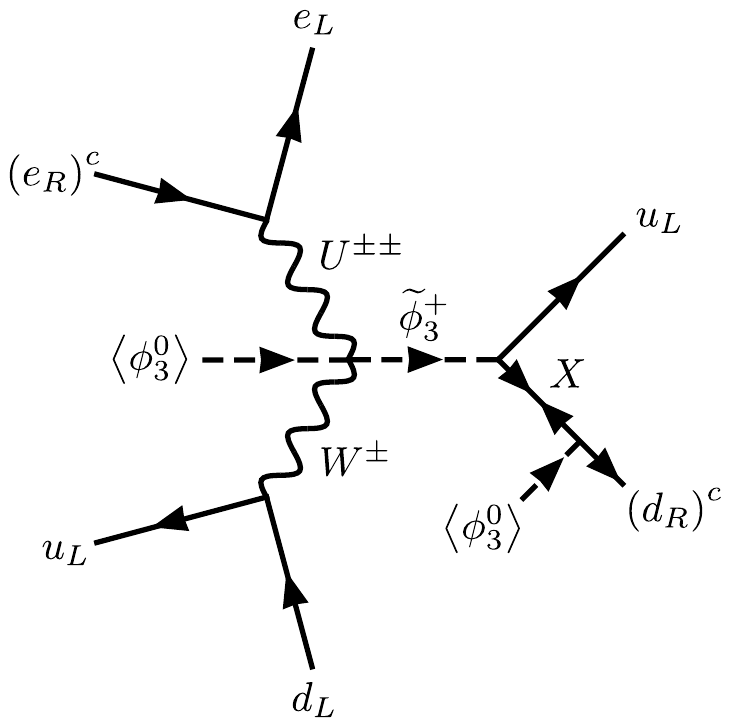}
			\par
		\end{centering}
		\protect\caption{\label{fig:SRT2}
			Short-range contributions to $\znbb$ decay in the PPF model (with a 331 gauge group). 
			To the left: subdominant contributions via gauge boson mixing. To 
			the right: diagram obtained if one extends the model with a new vector fermion $X$ --- see text.}
	\end{figure}
	\par
\end{center}

Before closing this section, we will also discuss possible short-range 
contributions to $\znbb$ decay in 331 models, which are shown in figure \ref{fig:SRT2}. 
Consider first the diagram to the left: it exists in any 
variant of the PPF model in which $V^+$ and $W^+$ mix.
One can estimate the constraint on the mixing angle $\theta_{VW}$ imposed by this 
diagram:
\begin{equation}\label{eq:SRS}
\Big(\frac{g_L^4}{8}\frac{p_F}{m_W^4m_{U^{++}}^2}\theta_{VW} \Big)
/\Big(\frac{G_F^2}{2 m_P}\Big) = C_3^{LL} \lsim 1.1 \times 10^{-8},
\end{equation} 
where the numerical constraint is taken from \cite{Gonzalez:2015ady}
using the limit from $^{136}$Xe \cite{KamLAND-Zen:2016pfg}. Taking
$m_{U^{++}} = 1$ TeV, this results in $\theta_{VW} \lsim 1.6 \times
10^{-2}$, several orders of magnitude weaker than the constraint
derived from the long-range diagram. This result is not surprising,
since the short-range diagram suffers from two suppression factors:
(a) as mentioned already in the introduction, short-range matrix
elements are smaller than long-range ones; and (b) relative to the
long-range part there is an additional suppression factor of $p_F
m_W/m_{U^{++}}^2$.  Similar arguments also render unimportant those diagrams
in which the $U^{++}$ is replaced by the doubly-charged
scalar ${\widetilde \phi}_1^{++}$ \cite{Montero:2000ar}. (In this case,
the suppression factor $p_F m_W/m_{U^{++}}^2$ has to be replaced by
a factor of roughly $m_e/m_{{\tilde \phi}_1^{++}}$, leading to
the same conclusions.)

More important short-range contributions can exist in extended 331 
models; the diagram on the right of figure \ref{fig:SRT2} is one such example. The 
idea behind this extension is to generate lepton number violation 
via a Yukawa coupling of $\phi_3$ to an exotic vector-like quark 
$X$. In this case $X=\left(\overline{\mathbf{3}},\mathbf{2},1-5/6\right)
$ is an exotic vector-like doublet. 
Its component with electric charge $-1/3$ can mix with the down-type quarks of the 
standard model, as indicated in the figure. At the $SU(3)_L$ level 
the $X$ can be embedded into the representation $\widetilde{X}=\left(\overline{\mathbf{3}},\overline{\mathbf{3}},1/3\right)$,
in which case the following new Lagrangian terms are allowed by the symmetries of the model:
\begin{eqnarray}
{\mathscr{L}}_X = Y_{QX} Q_{1,2} \widetilde{X} \phi_3^{*} 
+ Y_{dX}{\widetilde{X}^c}d^c\phi_3^{*} + m_X \widetilde{X}^c \widetilde{X} +\textrm{h.c.}\,,
\end{eqnarray}
assuming that the vector partner of $\widetilde{X}$, $\widetilde{X}^c$, is also present. This construction makes it possible to generate the $d=11$ operator 
${\cal O}_{54} = Q Q {\overline Q}d^c L {\overline e^c}HH$. The following is a rough 
estimate of the size of this diagram:
\begin{equation}
\Big(\frac{g_L^4}{8}\frac{\left\langle \phi_{3}^{0}\right\rangle ^{2}}{m_W^2m_{U^{++}}^2m_{\phi_3}^2m_X}Y_{QX}Y_{dX}\Big)
/\Big(\frac{G_F^2}{2 m_P}\Big) = C_5^{LL} \lsim 9.5 \times 10^{-9},
\end{equation}
where the numerical constraint is taken again from
\cite{Gonzalez:2015ady}. For $m_{U^{++}}=m_{\phi_3}=m_X\equiv\Lambda$ 
we estimate $\Lambda \gsim 2.7 (Y_{QX}Y_{dX})^{1/5}$ TeV. Note that 
this limit is of the order of what is expected for a $d=9$ operator, 
despite being a $d=11$ formally. This can be understood easily, 
since the diagram contains an ordinary $W_L$ boson.

\subsection{Models based on the $SU(4)\times SU(2)_L\times SU(2)_R$ group}
\label{subsect:422}

The exotic vector representation $T \equiv \left(\mathbf{3},\mathbf{1},\frac{2}{3}\right)$
was earlier found, in section \ref{sec:2}, to be potentially
relevant for neutrinoless double beta decay.
And it turns out that this field is contained in the adjoint representation of the $SU(4) \times SU(2)_L$ group. However, realistic models incorporating the Standard Model fermions require an extra $U(1)_R$, which can be seen as a remnant subgroup of an $SU(2)_R$ (the full group then becomes the so-called Pati-Salam group). We shall consider here this last scenario where on top of the $T$ vector field, the $W_R$ gauge bosons are also present.
The discussion will be brief for two reasons: (a) the phenomenology 
is very similar to the one of the left-right symmetric model 
discussed above, and (b) compared to other lower limits on the 
Pati-Salam scale, $\znbb$ decay provides very weak constraints. 

The main features of models based on the $SU(4)\times SU(2)_L\times SU(2)_R$ group are the following. Left-handed
quarks and leptons form a $\left(\mathbf{4},\mathbf{2},\mathbf{1}\right)$ representation, while the standard model right-handed fermions are part of a $\left(\overline{\mathbf{4}},\mathbf{1},\mathbf{2}\right)$. Note that this
last representation also contains right-handed neutrinos $(\nu_R)^c$.
Breaking of the Pati-Salam group to the SM one can be done either directly or via an intermediate $SU(3)_C\times SU(2)_L\times SU(2)_R\times U(1)_{B-L}$ symmetry group. In the latter case, the discussion of $\znbb$ decay becomes very similar
to the one presented in section \ref{subsect:LR}. In particular, lepton 
number violation is generated by the spontaneous breaking of 
$SU(2)_R \times U(1)_{B-L}$, which we assume is achieved with  
a non-zero vacuum expectation value of a $\Delta_R$ scalar field.

\begin{center}
	\begin{figure}[tbph]
		\begin{centering}
			\includegraphics[scale=0.77]{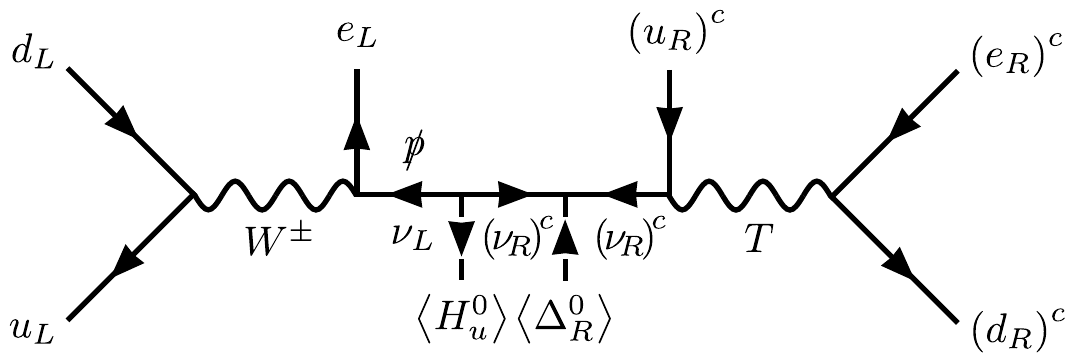}
			\\ \vskip5mm
			\includegraphics[scale=0.77]{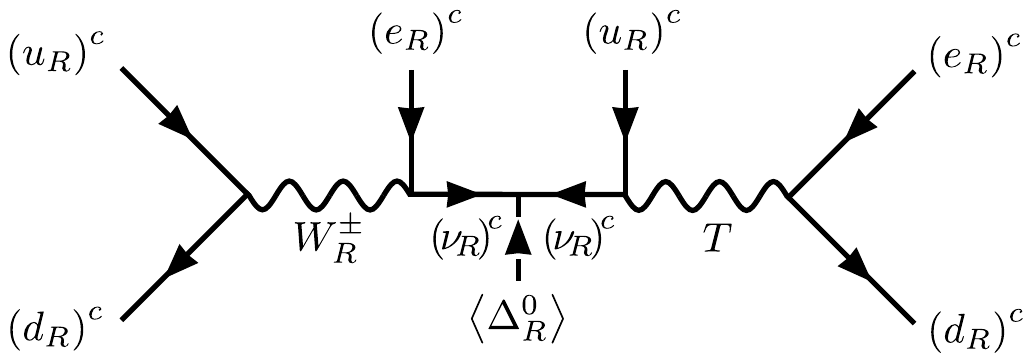}
			\includegraphics[scale=0.77]{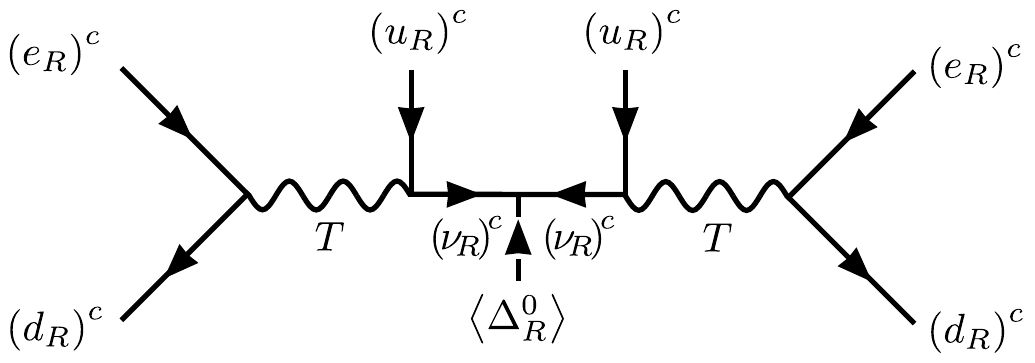}
			\par
		\end{centering}
		\protect\caption{\label{fig:PSdiags}
			Some neutrinoless double beta decay diagrams involving vectors fields which are present in models based on the Pati-Salam group. 
			The diagram on top shows a long-range contribution, while the two on the bottom illustrate short-range contributions.}
	\end{figure}
	\par
\end{center}

Figure \ref{fig:PSdiags} shows $\znbb$ decay diagrams containing the
vector $T$. A comparison with figure \ref{fig:LRdiags} reveals that
these diagrams are obtained by replacing the $W_R$'s with $T$'s, with
the matching replacements of lepton and quark fields. (Of course,
the coupling constant $g_R$ has to be replaced by $g_4$ as well.)  In all
cases, the source of lepton number violation is the Majorana mass term
for $(\nu_R)^c$. The diagram at the top of the figure is long-range
and it is proportional to $\pslash$ but at the same time it is
suppressed by light-heavy neutrino mixing. On the other hand, the two
diagrams at the bottom of figure \ref{fig:PSdiags}
are both short range diagrams which
provide lower limits on $\frac{g_{R}^2g_4^2}{m_{W_R}^2m_T^2\langle
	m_N\rangle}$ and/or $\frac{g_4^4}{m_T^4\langle m_N\rangle}$. As
in section \ref{subsect:LR}, the
numerical limits will again be of the order of $1-2$ TeV.

Since the gauge bosons couple to all fermion generations 
with the same strength in the Pati-Salam model, very stringent lower limits on the 
scale of the Pati-Salam group breaking can be derived from lepton \textit{flavor} 
violating meson decays. Naive limits are of the order of 
several $100$'s of TeV, but fermion mixing makes it possible to 
suppress certain decays, resulting in much reduced bounds 
on the Pati-Salam scale. Note that, because leptons and quarks 
are members of the same multiplet of the Pati-Salam group, besides the 
CKM and PMNS matrices which regulate interactions of the $W^\pm$ boson with fermions, there are also analogous matrices describing fermion mixing under $T$ interactions. However,  
as discussed in \cite{Kuznetsov:2012ai}, 
not all meson decays can be simultaneously suppressed 
by mixing, hence an unavoidable bound of $m_T \gsim 38$ TeV 
can be derived, which is significantly more stringent than the 
bounds from $\znbb$ decay.

\section{\label{sec:4}Conclusions}

We have systematically studied contributions of vector fields to
neutrinoless double beta decay.
First, we have identified all possible exotic vector representations which can
participate in $d=9$ and $d=11$ $\znbb$ decay
operators at tree-level.  It turns out that there are 46 possibilities. Then, in a
second step, we have searched for the minimal gauge groups for which
the vectors in our list are contained in the adjoint representation of
these extended gauge groups. Nevertheless, such a search does not take into account two important facts.
Firstly, some vectors cause proton decay and thus require a very high scale of
symmetry breaking, which in turn suppresses the $\znbb$ decay rate to negligible values.
But more importantly,
most of the relevant vector representations require non-standard embeddings of
the standard model gauge group, making it impossible to construct viable models
with the correct fermion spectrum.

The phenomenology associated to the few remaining vectors and groups was then discussed
in more detail. The valid groups are the left-right
symmetric group, the $SU(3)_C\times SU(3)_L\times U(1)_X$ (331) group
and the Pati-Salam group. The first one has been known to give
potentially important contributions to $\znbb$ decay for a long time,
so we only briefly discussed this case. For 331-based models, we have entered in
some more detail, discussing possible long- and short-range
contributions to $\znbb$ decay. We have placed a strong emphasis on the
identification of lepton number (and its violation) in these models, in order to 
be sure that the correct interpretation of $\znbb$ decay bounds was
found. In the case of the Pati-Salam group, we identified some new
contributions to $\znbb$ decay which are expected to be sub-dominant.
In other words, other constraints on the
scale of the Pati-Salam group are much more stringent than the ones
derived from $\znbb$ decay. Finally we remarked that constraints on
the mass and interactions of any vector in our list which is not a gauge boson can,
of course, also be derived from $\znbb$ decay.

\section*{\vspace{-0.15cm}Acknowledgments}

We thank Luca Di Luzio and Avelino Vicente for comments and discussions. This work was supported by the Spanish grants FPA2014--58183--P,
Multidark CSD2009--00064 and SEV--2014--0398 (from the
\textit{Ministerio de Economía, Industria y Competitividad}), as well
as PROMETEOII/2014/084 (from the \textit{Generalitat Valenciana}). RF
was also funded by the Juan de la Cierva-formación grant
FJCI--2014--21651 (from the Spanish \textit{Ministerio de Economía,
	Industria y Competitividad}).

\end{document}